\documentclass{elsart}
\usepackage{graphicx}

\begin{document}

\begin{frontmatter} 

\title{Mean-field model for the interference
of matter waves from a
three-dimensional optical trap}
 
\author{Sadhan K. Adhikari$^1$ and Paulsamy Muruganandam$^{1,2}$}

\address{$^1$Instituto de F\'{\i}sica Te\'orica, Universidade Estadual
Paulista,\\
01.405-900 S\~ao Paulo, S\~ao Paulo, Brazil\\
$^2$Centre for Nonlinear
Dynamics, Department of Physics, \\ Bharathidasan University,
 Tiruchirapalli
620 024, Tamil Nadu, India}

\date{\today}
\maketitle

\begin{abstract}

Using the mean-field time-dependent Gross-Pitaevskii equation we study the
formation of a repulsive Bose-Einstein condensate on a combined optical
and harmonic traps in two and three dimensions and subsequent generation
of the interference pattern upon the removal of the combined traps as in
the
experiment by Greiner {\it et al.} [Nature (London) {\bf 415}, 39 (2002)].
For optical traps of moderate strength, interference pattern of 27 (9)
prominent bright spots is found to be formed in three (two)  dimensions on
a cubic (square) lattice in agreement with experiment. 
Similar interference pattern can also be formed upon removal of
the optical lattice trap only. The pattern so formed can oscillate for a
long time in the harmonic trap which can be observed experimentally.

\end{abstract}
  
\begin{keyword}
Bose-Einstein condensation, Interference of matter wave
\PACS{03.75.-b, 03.75.Lm, 03.75.Kk}
\end{keyword}
 
\end{frontmatter}

The recent experimental observation of trapped Bose-Einstein condensates
(BECs) in alkali-metal atoms \cite{exp} has offered new possibility for
the
realization of matter-wave interference in the laboratory \cite{kett1}.
The major breakthrough in this endeavor has come from the formation of
harmonically trapped BEC on optical-lattice periodic potential in one
\cite{1,2} and three \cite{greiner,stoof} dimensions which has permitted
the
study of quantum phase effects on a macroscopic scale such as interference
of matter waves.  Innovative interesting
experiments on interference of matter waves from different optical lattice
centers have been performed. The phase coherence between
different sites of a trapped BEC on an optical lattice has been
established in two recent experiments \cite{1,2} through the formation of
distinct interference pattern when the traps are removed.

Among these experiments on BEC in optical lattice potential, the one worth
mentioning is the observation by Cataliotti {\it et al.} \cite{cata} of a
robust pattern in the form of three bright spots generated via the
interference of matter waves coming from different optical lattice centers
in one dimension after the removal of the traps. The recent
three-dimensional extension of this experiment by Greiner {\it et al.}
\cite{greiner} is a landmark of all such attempts \cite{stoof}.  The
observation of a sharp interference pattern of 27 prominent bright spots
arranged on a cubic lattice in that experiment upon removal of the
combined traps from a BEC formed in three-dimensional optical and harmonic
traps is a clear manifestation of phase coherence over the initial
condensate. There have been several theoretical studies on different
aspects of a BEC in a optical lattice potential \cite{th}.

In the recent experiment, Greiner {\it et al.} \cite{greiner} provided
quantitative measurement of the formation and evolution of interference
pattern upon free expansion of a harmonically trapped BEC of repulsive
$^{87}$Rb atoms formed on an optical lattice potential after removing both
traps. They also continued their investigation by increasing gradually the
strength of the optical lattice potential using a standing-wave laser beam
of increased intensity and found that the phase coherence between
different sites is lost. Consequently, the interference pattern is
destroyed for stronger optical lattice potential. This phenomenon is
termed \cite{greiner} the superfluid to Mott insulator transition. The
superfluid state is the one with phase coherence at different optical
lattice sites and the insulator state is the one where the phase coherence
is destroyed by a strong optical lattice potential. In the superfluid
phase atoms can move from one optical lattice site to another by quantum
tunneling, whereas in the Mott insulator phase this tunneling is stopped.
The motion of atoms in the two phases are similar to that of electrons in
semi-conductors and insulators and hence the name Mott insulator to one of
the phases.  In this Letter we investigate how well the superfluid state
on the three-dimensional optical lattice as well as the formation of the
prominent interference pattern upon free expansion can be described by the
mean-field nonlinear time-dependent Gross-Pitaevskii (GP) equation
\cite{8}.

The time-dependent BEC wave function $\Psi({\bf r};\tau)$ at position
${\bf r}$ and time $\tau $ is described by the following mean-field
nonlinear GP equation \cite{8} \begin{eqnarray}\label{a} \left[-
i\hbar\frac{\partial }{\partial \tau} -\frac{\hbar^2\nabla^2 }{2m} +
V({\bf r}) + gN|\Psi({\bf r};\tau)|^2 \right]\Psi({\bf r};\tau)=0,
\end{eqnarray} where $m$ is the mass and $N$ the number of atoms in the
condensate, $g=4\pi \hbar^2 a/m $ the strength of interatomic interaction
with $a$ the atomic scattering length.  The combined three-dimensional
harmonic and optical lattice traps are given by $ V({\bf r}) =\frac{1}{2}m
\omega ^2( x^2+ y^2+ z^2) +V_{\mbox{opt}}$ where ${\bf r}\equiv ( x, y,
z)$ with $-\infty > x, y, z>\infty$, where
  $\omega$ is the angular
frequency of the
harmonic trap
 and $V_{\mbox{opt}}$ is
the optical lattice trap introduced later.  
The normalization condition  is
$ \int d{\bf r} |\Psi({\bf r};\tau)|^2 = 1. $

In three dimensions, the wave function can be written as $\Psi({\bf r},
\tau)= \psi( x, y,z,\tau )$. Now transforming to dimensionless variables
$x_1 =\sqrt 2 x /l$, $x_2=\sqrt 2 y/l$, $x_3=\sqrt 2 z/l$, $t=\tau \omega,
$ $l\equiv \sqrt {\hbar/(m\omega)}$, and ${ \varphi(x_1,x_2,x_3;t)} \equiv
\sqrt{{l^3}/{\sqrt 8}}\psi( x, y, z;\tau),$ Eq. (\ref{a}) becomes \cite{9}
\begin{eqnarray}\label{d1}  
\biggr[&-& i\frac{\partial }{\partial t}
-\sum_{i=1}^3\frac{\partial^2}{\partial x_i^2} +\frac{1}{4}\sum_{i=1}^3
x_i^2 +\frac{V_{\mbox{opt}}}{\hbar \omega}\nonumber \\ &+& 8\sqrt 2 \pi
n\left|
{\varphi({x_1,x_2,x_3};t)}\right|^2 \biggr]\varphi({ x_1,x_2,x_3};t)=0,
\end{eqnarray} with nonlinearity $ n = N a /l$ and
normalization \begin{eqnarray}
\int
_{-\infty}^\infty dx_1 \int _{-\infty}^\infty dx_2\int
_{-\infty}^\infty 
dx_3  |\varphi({x_1,x_2,x_3};t)|^2 =1.\end{eqnarray}

The optical potential created with the standing-wave laser field of
wavelength $\lambda$ is given by $V_{\mbox{opt}}=V_0E_R\sum _{i=1}^3\sin^2
(k_Lx_i)$, with $E_R=\hbar^2k_L^2/(2m)$, $k_L=2\pi/\lambda$, and $V_0$ the
dimensionless strength of the optical potential \cite{greiner}. The
present dimensionless length unit is $l/\sqrt 2$ and time unit is $\omega
^{-1}$. In a typical experiment
$l/\sqrt 2\sim 1$ $\mu$m and  $\omega ^{-1}\sim 1$ ms.  
In terms of the
dimensionless laser wave
length $\lambda _0= \sqrt2\lambda/l $ and  the dimensionless
standing-wave energy parameter $E_R/(\hbar \omega)= 4\pi^2/\lambda
_0^2$, $ V_{\mbox{opt}}$ is given by 
\begin{equation}\label{pot}
\frac{ V_{\mbox{opt}}}{\hbar \omega}=V_0\frac{4\pi^2}{\lambda_0^2} 
\sum_{i=1}^3 \sin ^2 \left(
\frac{2\pi}{\lambda_0}x_i
\right).
\end{equation}

In the  experiment of Greiner  {\it et al.} \cite{greiner}
with repulsive $^{87}$Rb atoms in the hyperfine state $F=2,
m_F=2$,  $ \omega =
2\pi \times 24 $ Hz,
$\lambda=852$ nm, 
$m=1.441 \times 10^{-25}$ kg,  
$l=\sqrt {\hbar/(m\omega)} = 2.204$ $\mu$m,  $\omega ^{-1} =
1/(2\pi\times 24)=6.63$ ms and 
$\lambda _0= \sqrt2\lambda/l \simeq 0.547$. In that experiment 65 
lattice sites in the single direction are populated in the initial state
by  a total number of more than 500 000 atoms. Because of limitations in
computer processing time and memory we present here a  model
calculation with a smaller
condensate.

We solve   Eq. (\ref{d1}) numerically using a split-step
(four-step) time-iteration
method with the Crank-Nicholson discretization scheme described recently
\cite{11}.   The three kinetic
energy derivative terms  were treated in three separate
steps. All nonderivative terms are treated in the fourth step. 
 To calculate the initial state on optical lattice
we discretize the GP equation 
spanning $x_i$ from $-5$ to 5, $i=1,2,3$.  
The time iteration is started with the
known harmonic oscillator solution of   Eq. (\ref{d1}) for $n=0$. For a
typical condensate the chemical potential is much smaller than the typical
strength of the optical potential wells $E_R$, so that
passage of condensate atoms from one well to other can only proceed
through quantum tunneling.  The nonlinearity $n$ as well as the optical
lattice potential parameter $V_0$ are slowly increased by equal amounts in
$1000n$ steps of time iteration until the desired value of nonlinearity
and optical lattice potentials are attained. Then, without changing any
parameter, the solution so obtained is iterated several thousand times so
that a stable solution is obtained independent of the initial input and
time and space steps.  The solution then corresponds to the bound BEC
under the joint action of the harmonic and optical traps. 
To study the
expansion of the condensate upon the removal of the traps the initial
wave
function is loaded at the center of a bigger lattice defined by $-15 <x_i
<15, i=1,2,3.$ 
The numerical calculation was performed with nonlinearity
$n=1$, $V_0=5$  and $\lambda_0=1$, although the general trend of the
results presented in this Letter is
independent of this
particular choice.

However, we begin our study with a simpler model in two dimensions
governed by the following dimensionless GP equation
\begin{eqnarray}\label{d2}  \biggr[-i\frac{\partial }{\partial t}
-\sum_{i=1}^2\frac{\partial^2}{\partial x_i^2} +\frac{1}{4}\sum_{i=1}^2
x_i^2 +\frac{V_{\mbox{opt}}}{\hbar \omega} +
               n\left| {\varphi({x_1,x_2};t)}\right|^2
 \biggr]\varphi({ x_1,x_2};t)=0, \end{eqnarray} with nonlinearity $n$ and
normalization $\int _{-\infty}^\infty dx_1 \int _{-\infty}^\infty dx_2
|\varphi({x_1,x_2};t)|^2 =1.$ The sumation in the optical potential
(\ref{pot}) in this case is limited to two terms only. Although, the
results of this study in two dimensions cannot be directly compared with
experiment, this serves as an excellent illustration of the physics
involved in terms of this simpler model which makes the computation
easier.

In two dimensions the wave function of the initial condensate was
calculated by solving  Eq. (\ref{d2}) using a split-step (three-step) time
iteration
method  with the Crank-Nicholson scheme
\cite{11}
on a lattice spanning $x_i (i=1,2)$ from $-10$ to 10. 
The kinetic energy derivative terms
in
two directions are treated in two different steps and all other
nonderivative terms are treated in the third step.  The expansion of the
condensate was
studied by loading the initial wave function on a  larger lattice spanning
$x_i (i=1,2)$ from $-50$ to 50. 
In the
numerical calculation we used $\lambda_0 =1$, $V_0=5$,   $n=100$ in
Eq. (\ref{d2}).

\begin{figure}[!ht]
 
\begin{center}
 
\includegraphics[width=.5\linewidth]{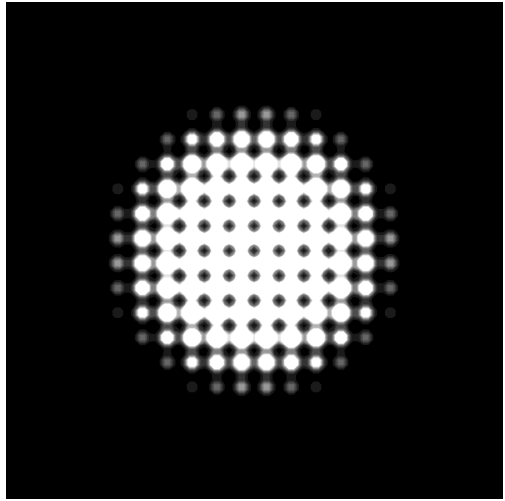}
\end{center}
 
\caption{The contour plot of the initial wave function in two dimensions 
on a $10 \times 10$
mat ($-5<x_i<5, i=1,2 $  )
for the ground-state BEC with $n=100$ and $V_0=5$. 
} 

\end{figure}

The projection of the initial wave function on the $x-y$ plane (contour
plot) in two dimensions 
is exhibited in Fig. 1.
  The bright positions are  the maxima of the wave function and the
dark regions represent minima. This figure clearly shows the formation of 
bright structures in the $x-y$ plane in the different optical potential
wells. About 16 lattice sites are populated in a single direction in the
initial state.

We next load the initial BEC of Fig.  1 at the center of a larger mat
(area $100\times 100$) and remove both the harmonic and the optical
lattice traps. The contour plot of the time evolution of the condensate is
shown in Fig. 2.  The initial wave function at $t=0$ is the same as the
one in  Fig. 1. However, the optical lattice pattern at $t=0$ is not
visible in Fig.  2 because of the (large) size of the mat in this plot.  
When the condensate is released from the combined traps, a matter-wave
interference pattern is formed in 0.25 time units.  The atom cloud
released from one lattice site expands, overlaps and interferes with atom
clouds from neighboring sites to form the robust interference pattern due
to phase coherence. No interference pattern can be formed without phase
coherence.  The interference pattern is composed of 9 prominent bright
spots arranged on a square lattice with 4 spots at corners, 4 at the
middle points of the sides and 1 at the center of the square. However,
there are some weaker secondary spots in the interference pattern. The
interference pattern keeps on expanding at larger times as shown in Fig. 
2.  Similar interference pattern was observed \cite{cata}
also in one dimension, where
 the  pattern is composed
of three bright spots on a line   in the form  of a central peak 
 and two symmetrically spaced peaks, each containing about $10\%$ of 
total number of atoms, 
moving apart from the central
peak  with
a constant velocity.

\begin{figure}%[!ht]
 
\begin{center}
\includegraphics[width=0.8\linewidth]{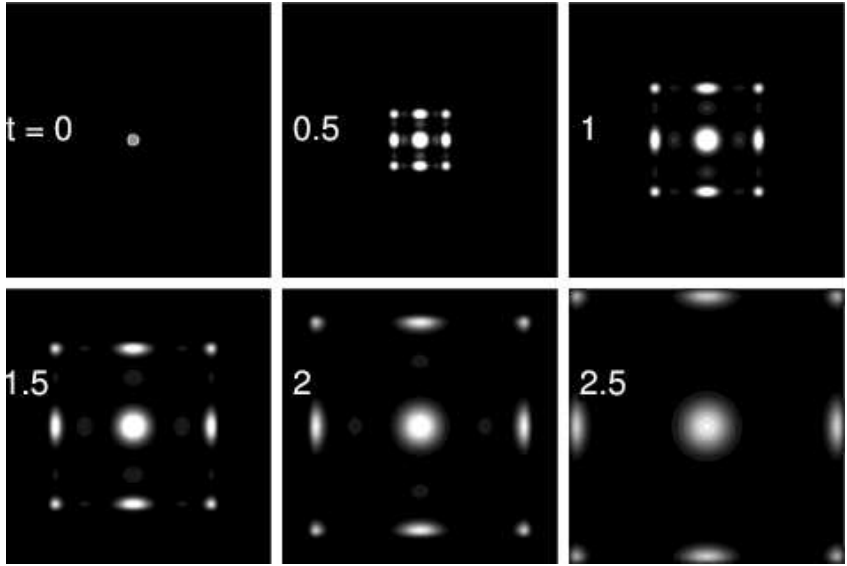}
\end{center}
 
\caption{Time evolution of the BEC of Fig.
 1 upon removal of both
optical
lattice
and harmonic traps at time $t=0$ on a $100\times 100$   mat ($-50<x_i 
<50, i=1,2$) showing the formation and evolution of the interference
pattern.}
\end{figure}

\begin{figure}[!ht]
 
\begin{center}
\includegraphics[width=0.8\linewidth]{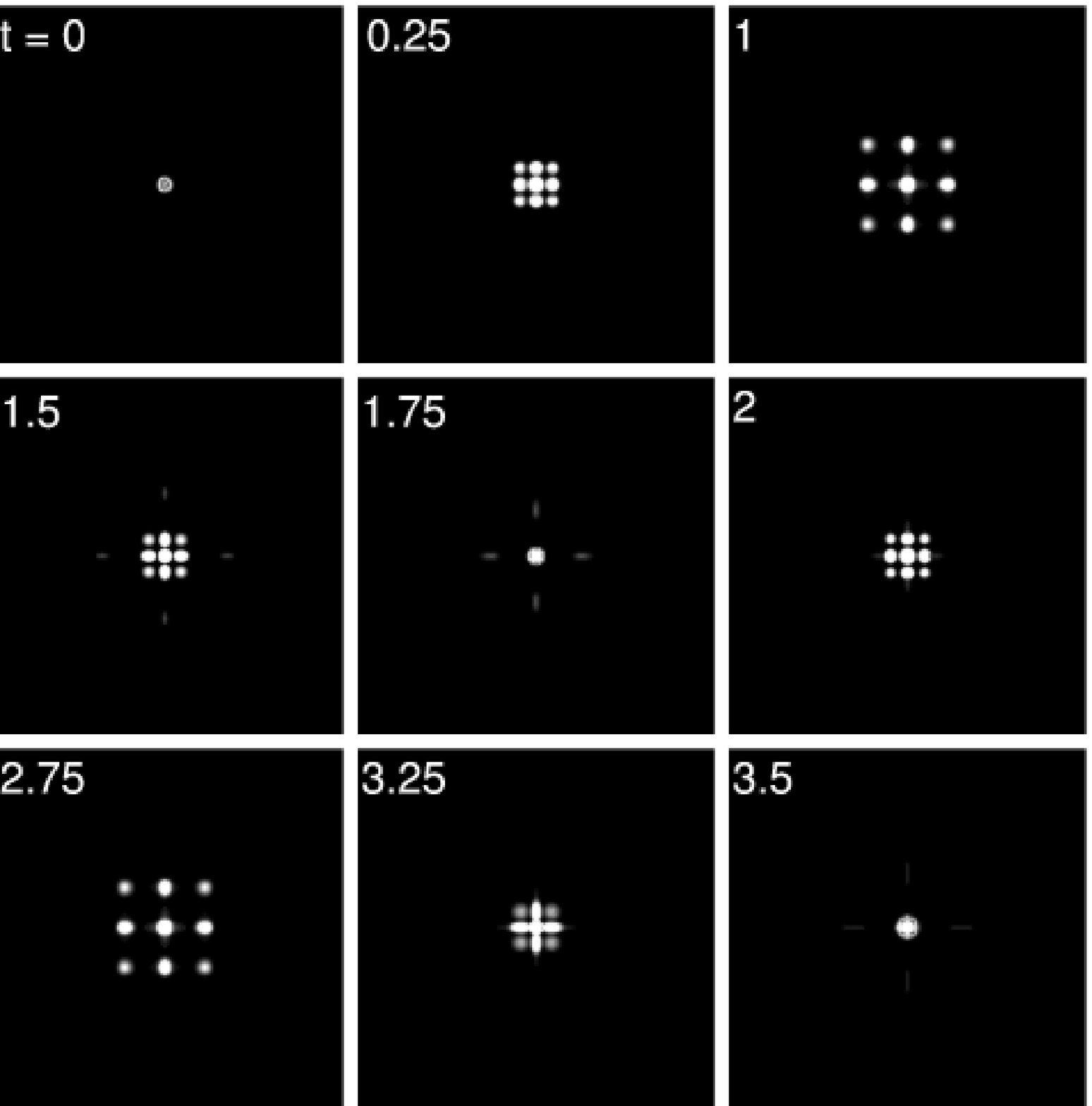}
\end{center}

\caption{Time evolution of the BEC of Fig. 1 upon removal of only the 
optical lattice trap at time $t=0$ on a $100\times 100$ mat 
($-50<x_i<50, i=1,2$) demonstrating the oscillating robust interference
pattern 
in the harmonic trap.}
 
\end{figure}

Usually a repulsive condensate without a confining trap should 
disparse quickly \cite{ad}. The robust interference pattern formed from a
repulsive condensate with very little spreading is similar to bright
solitons.  
True bright solitons can be formed in attractive BEC only and have been 
observed  experimentally 
\cite{sol,sol1}. 
The phase coherence between 
the components of BEC at a large number of  sites of optical lattice is
responsible
for the generation of the interference pattern with very little
spreading. Without the initial phase
coherence a repulsive condensate in the absence of a trap will disappear
in  few units of time \cite{ad}. Each of the moving interference peaks 
are similar to atom laser \cite{1,al} which can be used in the scattering
of
two 
coherent BECs and other purposes.

To study the robustness of the interference pattern we remove only
the optical trap at time $t=0$ and allow the condensate to evolve in the
harmonic trap alone so that the interference pattern cannot escape to
infinity. In this case the interference pattern is formed quickly which
tends to expand at small times. However, because of the confining harmonic
trap, after some initial expansion the interference pattern starts to
shrink towards the center. Eventually, it shrinks to a central spot and
starts to expand again.  This expansion and shrinking of the interference
pattern without considerable distortion is repeated over many cycles of
which we show the two first cycles in Fig. 3. The period of this
oscillation is 3.5 units of time. The largest size of the interference
pattern occurs approximately at $t=1$ and 2.75. This periodic oscillation
of the interference pattern can be observed experimentally and the present
mean-field prediction could be compared with future experiments.

\begin{figure}%[!ht]
 
\begin{center}
\includegraphics[width=0.8\linewidth]{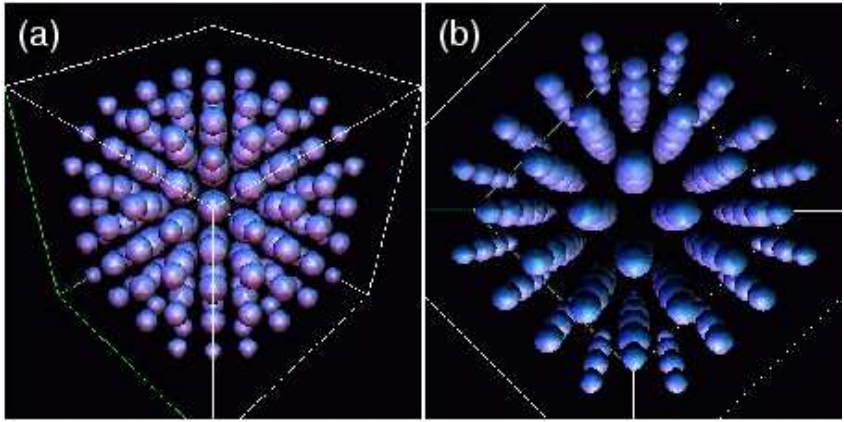}
\end{center}

\caption{The three-dimensional contour plot of the initial wave function
on a cubic lattice  
of size  $3 \times 3 \times 3$
 ($-1.5<x_i<1.5, i=1,2,3 $  )
for the ground-state BEC with $n=1,$ $\lambda _0 =1$ and $V_0=5$: (a) view 
along a diagonal and    (b) along one of the axes of the cube. 
} 
 
\end{figure}

\begin{figure}[!ht]
 
\begin{center}
\includegraphics[width=0.8\linewidth]{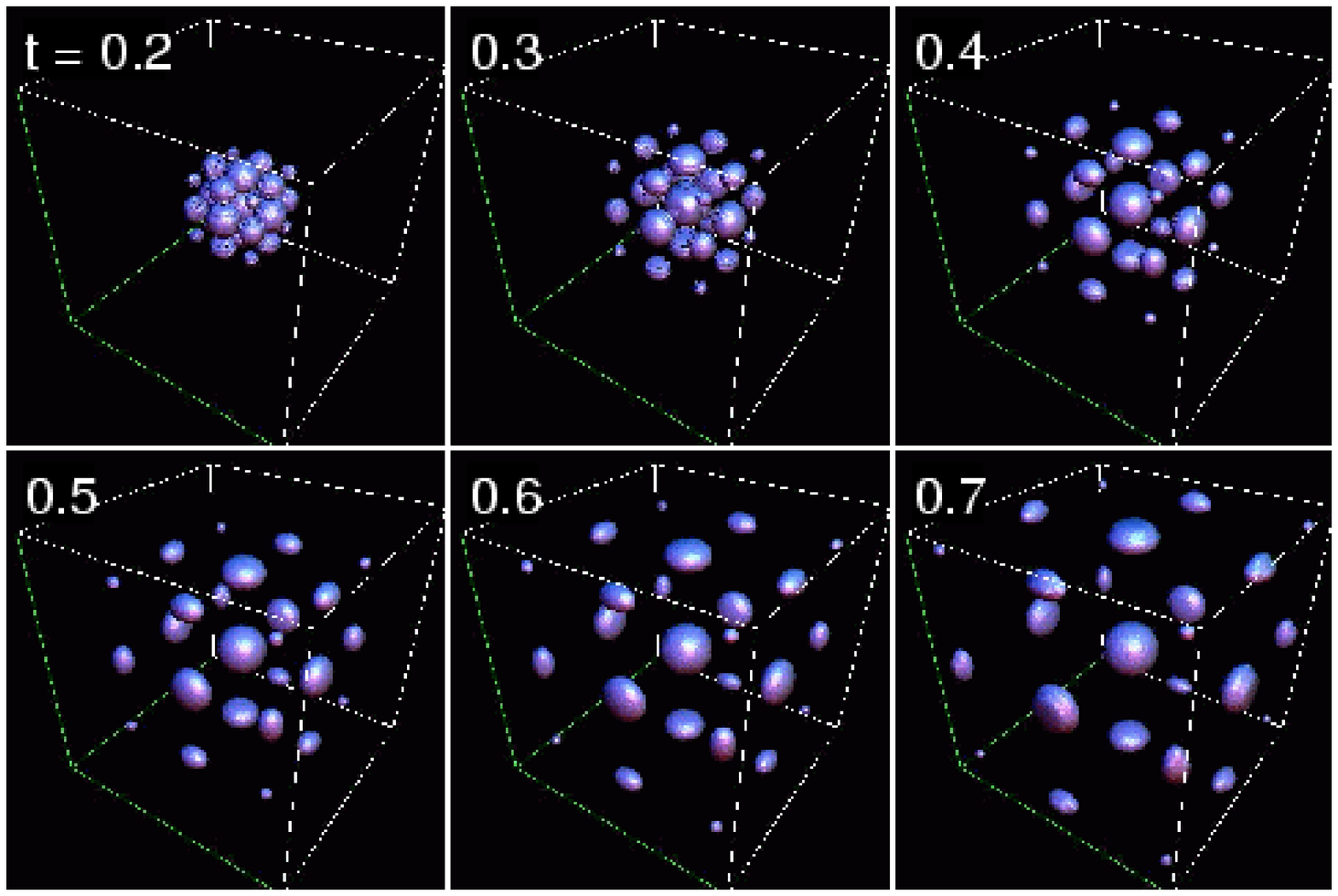}
\end{center}

\caption{Time evolution of the BEC of Fig. 4 upon removal of both
optical
lattice and harmonic traps at time $t=0$ on a cubic lattice of size
$30\times 30\times 30$ ($-15<x_i <15, i=1,2,3$) showing the formation and
evolution of the interference pattern.}

\end{figure}

After this preliminary study in two dimensions we consider the formation
of a BEC in a three-dimensional optical lattice trap and the subsequent
formation
of the interference pattern upon simultaneous removal of the optical
lattice and harmonic traps. The initial solution for $n=1$, $V_0 =5$ and 
$\lambda_0 =1$ was calculated on a cube of size
$10 \times 10 \times 10$. In Fig. 4
we show the contour plot of the central part of the initial  wave function
on a cube of
size $3 \times 3
\times 3$. 
Two views of the three-dimensional pattern is shown. In Fig. 4 (a) we
show the view in the diagonal  direction and in Fig. 4 (b) we  show that
along
one of the axes of the cube. The formation of the condensate in the
three-dimensional cartesian lattice is clearly demonstrated in
Fig.  4. About 8 to 10 lattice sites are populated in a single direction
in
the
initial state of which the central portion is shown in Fig. 4.

Next to study the formation and evolution of interference pattern in three
dimensions we load the bound state of Fig.  4 at the center of a cube of
size $30 \times 30 \times 30$ and remove the combined harmonic and optical
lattice traps at $t=0$.  In Fig.  5 we show the time evolution of the
condensate. Upon the removal of the traps a matter-wave interference
pattern is formed quickly in less than 0.2 units of time. This pattern is
composed of 27 prominent bright spots arranged on a cubic lattice: 8 at
the corners, 12 at the middle points of the sides, 6 at the center of each
of the bases, and 1 at the center of the cube. As in one and two
dimensions, this pattern is also very robust and expands without
considerable distortion as shown in Fig.  5. In both two and three
dimensions the structure and the shape of the interference pattern is
independent of the number of lattice sites occupied in the initial state,
the nonlinearity $n$, or the strength  $V_0$ and the wave length
$\lambda_0$
of the optical trap potential.

In conclusion, to understand theoretically the experiment by Greiner {\it
et al.} \cite{cata}, we have studied in detail the phase coherence in a
cubic condensate loaded in a combined harmonic and optical lattice traps
using the solution of the full GP equation in three dimensions. We also
performed a similar study on a two dimensional condensate. Using the
split-step Crank-Nicholson method, first we obtain the initial wave
function of the condensate in the combined harmonic and optical lattice
traps. Then we study the time evolution of the system after the removal of
both the traps. Robust interference patterns of 9 (in two dimensions)  
and 27 (in three dimensions) prominent bright spots are formed upon the
removal of the traps.  The present robust interference pattern for
repulsive atoms with very little spreading  is similar to  bright
solitons in case of attractive
condensates.  True bright solitons can be formed  only in
attractive BEC \cite{sol,sol1}. 
The
formation of the interference pattern clearly demonstrates the
phase coherence in the initial condensate  on the optical lattice. 
Each of the moving interference peaks formed of coherent matter wave  
is similar to a atom laser observed experimentally 
\cite{1,al}.  
The interference pattern can also be created  by removing only the
optical lattice trap.  
On removal of the optical lattice trap the interference pattern 
is found to oscillate in the harmonic trap without much
distortion. This phenomenon demonstrates the robustness of the
interference pattern
and is studied in detail in two dimensions. This oscillation can be
observed experimentally and the result of future experiments can be
compared with the present mean-field prediction.

\ack 

The work was supported in part by the CNPq and FAPESP
of Brazil.

 \end{document}